
\magnification=1250
\overfullrule=0pt
\baselineskip=19pt

\centerline{\bf Phase Diagram of the Spin-One Heisenberg System}
\centerline{\bf with Dimerization and Frustration}
\vskip 0.2 in

\centerline{Swapan Pati$^{\ast}$, R. Chitra$^{\ast \ast}$, Diptiman
Sen$^{\dagger , +}$,}
\centerline{H. R. Krishnamurthy$^{\ast \ast , +}$ and S.
Ramasesha$^{\ast , +}$}

\centerline{\it $^{\ast}$ Solid State and Structural Chemistry Unit,}
\centerline{\it Indian Institute of Science, Bangalore 560012, India}

\centerline{\it $^{\ast \ast}$ Physics Department, Indian Institute of
Science,}
\centerline{\it Bangalore 560012, India}

\centerline{\it $^\dagger $ Centre for Theoretical Studies, Indian
Institute of Science,}
\centerline{\it Bangalore 560012, India}

\centerline{\it $^{+}$ Jawaharlal Nehru Centre for Advanced Scientific
Research,}
\centerline{\it Indian Institute of Science Campus, Bangalore 560012,
India}

\vskip 0.2in

\line{\bf Abstract \hfill}

We use the density matrix renormalization group method to study the ground
state properties of an antiferromagnetic spin-$1$ chain with a next-nearest
neighbor exchange $J_2 ~$ and an alternation $\delta$ of the nearest
neighbor exchanges. We find a line running from a gapless point at
$(J_2 , \delta) = (0, 0.25 \pm 0.01)$ upto an almost gapless point at
$(0.725 \pm 0.01, 0$ such that the open chain ground state is $4$-fold
degenerate below the line and is unique above it. A disorder line
$2 J_2 + \delta = 1$ runs from $\delta =0$ to about $\delta =0.136$.
To the left of this line, the peak in the structure factor $S(q)$ is
at $\pi$, while to the right of the line, it is at less than $\pi$.

\vskip 0.2in

\noindent
PACS numbers: 75.10.Jm, 75.50.Ee

\vfill
\eject

While the isotropic spin-$1/2$ Heisenberg antiferromagnetic chain
has been extensively studied using a variety of
analytical and numerical techniques [1], the corresponding spin-$1$ chain
has been studied in much less detail [2-4].
Interest in spin-$1$ chains grew after Haldane's conjecture that integer
spin chains with a nearest-neighbor (nn) exchange should have a gap while
half-integer spin chains should be gapless. This observation was based on
a non-linear sigma model (NLSM) field theory description of the low-energy
excitations [5]. The NLSM approach can be generalized to include other
features such as dimerization (an alternation $\delta$ of the nn
exchanges) and a next-nearest-neighbor (nnn) exchange $J_2 ~$ [6], and it
leads to interesting predictions. For
instance, the spin-$1$ model should exhibit a gapless point at some
critical value of $\delta$. If the nnn exchange is large enough, the spin
chain goes over from a Neel like "phase" [7] to a spiral "phase" and a
different kind of NLSM field theory becomes applicable [8,9].
This predicts a gap for {\it all} values of the spin.

In a recent paper [10], we studied the $J_2 - \delta$ model for a
spin-$1/2$ chain using the density matrix renormalization group (DMRG)
method [2,11]. In this Letter, we extend this study to the
spin-$1$ chain with both dimerization and frustration and compare our
results with the field theoretic expectations. The major surprise which
we discover is an almost gapless point at $(J_2 =0.725, ~\delta =0)$
which is contrary to the field theory expectation. We suggest that this
point may be close to a critical point which is described by a $SU(3)$
symmetric conformal field theory (CFT) [12,13].

We have studied both open and periodic chains with an even number of
sites governed by the Hamiltonian
$$H ~=~ \sum_i ~[ ~1 ~-~ (-1)^i ~\delta ~] ~ {\bf S}_i \cdot
{\bf S}_{i+1} ~ + ~J_2 ~\sum_i ~{\bf S}_i \cdot {\bf S}_{i+2} ~.
\eqno(1)$$
with the limits of $i$ being interpreted as appropriate.
We restrict our attention to the region
$J_2 ~\ge 0$ and $0 \le \delta \le 1$. We study various regions in the
$(J_2 , \delta)$ plane using DMRG.

The DMRG technique involves systematically building up the chain to a
desired number of sites starting from a very short chain by adding two sites
at a time. The initial chain of $2n$ sites (with $n$ a small enough integer)
is diagonalized exactly. The reduced density matrix for the left $n$ sites
is computed from the ground state of the $2n$ chain Hamiltonian by
integrating over the states of the right $n$ sites. This density matrix
is diagonalized and a matrix representation of the $n$-site Hamiltonian
is obtained in a truncated basis with $m$ basis vectors which are the
eigenvectors of the density matrix corresponding to its $m$ largest
eigenvalues. The Hamiltonian matrix for the $2n+2$ chain is then obtained
in the $(2s+1)^2 m^2 ~$ dimensional direct
product subspace constructed using
the truncated basis of the left and the right halves of the $2n$ chain and
the full space of the two additional spins which are inserted in the middle.
After obtaining the ground state of the $2n+2$ chain in the truncated basis,
the density matrix of half the chain, now with $n+1$ sites, is computed.
The procedure is repeated till one reaches the desired chain length $N$.
The accuracy of
the DMRG technique depends crucially on the number of eigenvalues of the
density matrix, $m$, which are retained. We worked with $m=100$ to $120$
over the entire $J_2 - \delta$ plane after checking that the DMRG results
obtained using these values of $m$ agree well with exact numerical
diagonalizations of chains upto $14$ sites. The chain lengths
we studied varied from $150$ sites for $J_2 > 0$ to
$200$ sites for $J_2 =0$. We tracked our results as a
function of $N$ to check that convergence had been reached well before
$150$ sites.

The "phase diagram" which we obtained is shown in Fig. 1. There is a solid
line marked $A$ which runs from $(0.25,0)$ to about
$(0.22 \pm 0.02, 0.20 \pm 0.02)$ shown by a cross. Within
our numerical accuracy, the gap is zero on this line and the correlation
length $\xi$ is as large as the system size $N$. The rest of the
phase diagram is gapped. However the gapped portion can be
divided into different regions characterized
by other interesting features. On the dotted lines marked $B$, the gap is
finite. Although $\xi$ goes through a maximum when we cross $B$ from either
side, its value is much smaller than $N$. There is a dashed line $C$
extending from $(0.65,0.05)$ to about $(0.725,0)$ on which the gap is very
small and $\xi$ is very large but not as large as $N$. Below the curve
$ABC$, the ground state for an {\it open} chain has a four-fold
degeneracy (consisting of $S=0$ and
$S=1$), whereas it is unique above the curve ($S=0)$. The dashed line marked
$D$ satisfies $2 J_2 + \delta = 1$, has an exactly dimerized ground state,
and extends from $(0,1)$ to about $(0.432,0.136)$. Below the curve $ABC$,
there is a line $E$ which goes down to about $(0.39,0)$.
Across $D$ and $E$, the position of
the peak in the structure factor decreases from $\pi$ (Neel) to less than
$\pi$ (spiral). (The positions of all the above points have an uncertainty
of $\pm 0.01$ unless otherwise stated). We will comment on all these
features of the phase diagram below.

For reasons explained below, the (almost) gapless point at $(0.725,0)$
is quite unexpected. So we have studied that point in more detail.
In Fig. 2, we present a plot of the gap versus $J_2 ~$ along the line
$\delta =0$. It is highly non-monotonic with a very small value at about
$J_2 =0.725$. Fig. 3 is a plot of the static structure factor $S(q)$
versus $q$ at four values of $J_2 ~$ near that point. (We studied open
chains with $150$ sites). For $J_2 =0.725 $ and
$0.735$, we see a pronounced peak at about $q_{max} = 112^o ~$. The peak
decreases in height and becomes broader as one moves away from those
two values of $J_2 ~$.
We estimate the maximum value of $\xi$ to be about $60$ sites. It also
decreases rapidly as we move away from those special values of $J_2 ~$.

It is natural to speculate that $(0.725,0)$ lies close to some critical
point which exists in a bigger parameter space. We believe that the
appropriate critical point may be the one discussed
in Refs. [12, 13]. Ref. 12 exactly solves a spin-$1$ chain which has nn
interactions of the form
$$H ~=~ \sum_i ~[ ~{\vec S}_i \cdot {\vec S}_{i+1} ~+~ (~{\vec S}_i \cdot
{\vec S}_{i+1} ~)^2 ~]~,
\eqno(2)$$
and finds that there are gapless modes at $q =0$ and $\pm ~120^o$.
This implies a peak in the structure factor at $q =120^o ~$ which is
not very far from the value we observe numerically. Ref. 13 argues that
the long-distance physics of this model is described by a CFT with
$SU(3)$ symmetry [14].

Briefly, the field theoretic analysis of spin chains with the inclusion
of $J_2 ~$ and $\delta$ proceeds as follows. In the
$S \rightarrow \infty$ limit, a classical treatment shows that the ground
state of the model is in the Neel phase (a colinear configuration) for
$4 J_2 + \delta^2 < 1$, and in  a coplanar spiral configuration for $4
J_2 + \delta^2 > 1$. To next order in $1/S$, one derives a
semiclassical field theory to describe the long-wavelength low-energy
excitations. The field theory in the Neel phase is the $O(3)$ NLSM with a
topological term [5,6]. The field variable is a unit vector $\vec \phi$
with the Lagrangian density
$${\cal L} ~=~ {1 \over {2 c g^2}} ~{\dot {\vec \phi}}^2 ~-~ {c \over {2
g^2}} ~{\vec \phi}^{\prime 2} ~+~ {\theta \over {4 \pi}} ~{\vec \phi}
\cdot {\vec \phi}^{\prime} \times {\dot {\vec \phi}} ~,
\eqno(3)$$
where $c = 2 S (1 - 4J_2 - \delta^2 ~)^{1/2} ~$ is the spin wave
velocity, $g^2 = 2 /[S (1 - 4 J_2 - \delta^2 ~)^{1/2} ]~$ is the coupling
constant, and $\theta = 2 \pi S (1 - \delta)$ is the coefficient of the
topological term. Note that $\theta$ is independent of $J_2 ~$ in the
NLSM. (Time
and space derivatives are denoted by a dot and a prime respectively). For
$\theta = \pi$ mod $2 \pi$ and $g^2 ~$ less than a critical value, the
system is gapless and is described by a CFT with an $SU(2)$ symmetry
[6,13]. For any other value of $\theta$,
the system is gapped. For $J_2 = \delta =0$, one therefore expects that
integer spin chains should have a gap while
half-integer spin chains should
be gapless. This is known to be true even for small values of $S$ like
$1/2$ (analytically) and $1$ (numerically) although the field theory is
only derived for large $S$. In the presence of dimerization, one expects
a gapless system at certain special values of $\delta$. For $S=1$, the
special value is predicted to be $\delta =0.5$. We see that the {\it
existence} of a gapless point is correctly predicted by the NLSM. However,
according to the DMRG results, its location is
at $\delta_c =0.25$ for $J_2 =0$ [3] and decreases with $J_2 ~$ as shown
in Fig. 1. These
deviations from field theory are probably due to higher order corrections
in $1/S$ which have not been studied analytically so far.

In the spiral phase, it is necessary to use a different NLSM which is
known for $\delta =0$ [8,9]. The field variable is now an $SO(3)$ matrix
$\underline R$ and the Lagrangian density is
$${\cal L} ~=~ {1 \over {2 c g^2}} ~{\rm tr} ~\Bigl( ~{\dot
{\underline R}^T} {\dot {\underline R}} ~P_0 ~\Bigr) ~-~
{c \over {2 g^2}} ~{\rm tr} ~\Bigl( ~
{\underline R}^{\prime T} {\underline R}^{\prime} ~P_1 ~\Bigr) ,
\eqno(4)$$
where $c = S (1 + y) {\sqrt {1 - y^2}} / y$, $g^2 = 2
{\sqrt {(1 + y) /(1 - y)}} /S$ with $1/y = 4 J_2 ~$, and
$P_0 ~$ and $P_1 ~$ are diagonal matrices with entries $(1,1,2 y (1 -
y) / (2 y^2 - 2 y + 1) )$ and $(1,1,0)$ respectively. Note that
there is no topological term; indeed, none is possible since $\Pi_2
(SO(3)) =0$ unlike $\Pi_2 (S^2) = Z$ for the NLSM in the Neel phase.
Hence there is no apparent difference between integer and half-integer
spin. A one-loop renormalization group [8] and large $N$ analysis [9]
then indicate that the system should have a gap for all values of $J_2 ~$
and $S$, and that there is no reason for a particularly small gap at any
special value of $J_2 ~$. (A similar conclusion is obtained from a bosonic
mean-field theory analysis of the frustrated spin chain [15]). The almost
gapless point at $J_2 =0.725$ for spin-$1$ is therefore surprising.

For $\delta < 0.25$ and $J_2 =0$, the spin-$1$ chain is known to exhibit
a `hidden' $Z_2 \times Z_2 ~$
symmetry breaking described by a non-local order parameter [3,16].
This leads to a four-fold degeneracy of the ground state for the open
chain. The degeneracy may be understood in terms of spin-$1/2$
degrees of freedom living at the ends of an open chain whose mutual
interaction decreases exponentially with the chain length [17]. We have
oberved this ground state degeneracy at all points below the curve
$ABC$ in Fig. 1, where the gap between the singlet and triplet states
vanishes exponentially with increasing chain length. Above the curve,
the ground state is unique. The situation is reminiscent of the $Z_2 \times
Z_2 ~$ symmetry breaking mentioned above. However, we have not yet directly
studied the non-local order parameter using DMRG.

We have also examined the structure factor $S(q)$. Since there is no
long-range order anywhere in the $J_2 - \delta$ plane (except for algebraic
order on the line $A$ in Fig. 1), $S(q)$ generally has a broad peak at
some $q_{max} ~$. To the left of lines
$D$ and $E$ in Fig. 1, $q_{max} ~$ is pinned at $\pi$, while to the right
of $D$ and $E$, $q_{max} ~< \pi$. Above the curve $ABC$, the cross-over
from the Neel to the spiral phase presumably occurs across the
straight line $2 J_2 + \delta =1$ (see below). Below $ABC$,
the cross-over has been determined purely numerically and seems to occur
across the line indicated as $E$ in Fig. 1. The
region of intersection between the cross-overs from Neel to spiral and
from four-fold degeneracy to a unique ground state is a small `hole' in
the phase diagram centred about the point $(0.435,0.12)$. Points in
this `hole' turned out to be extremely difficult to study using DMRG
because of poor convergence with increasing chain lengths.

The segment $D$ of the straight line $2 J_2 + \delta = 1$ indicated
in Fig. 1 can be shown to have a dimerized state as the exact ground
state. It is easy to show that a dimerized state of the form
$$\psi ~=~ [1,2] ~[3,4]~.... ~[N-1 ,N]~,
\eqno(5)$$
where $[i,j]$ denotes the normalized singlet combination of the spins on
sites $i$ and $j$, is an eigenstate of the Hamiltonian on that line.
To prove that (5) is the ground state, we decompose the Hamiltonian as
$$H ~=~ \sum_i ~H_i ~,
\eqno(6)$$
where each of the $H_i ~$ only acts on a cluster
of $n$ neighboring sites. Next, we numerically show
that (5) is a ground state of each of the $H_i ~$, and is therefore
a ground state of $H$ by the Rayleigh-Ritz variational principle. For
$n=3$, this proof that (5) is the ground state works
down to $\delta = 1/3$ [18].
Below that, (5) is no longer the ground state of any of the $3$-cluster
Hamiltonians $H_i ~$. But we can construct $4$-cluster
$H_i ~$ satisfying (6) such that (5) can be numerically shown to be a
ground state of each of those. This allows us to prove
that (5) is the ground state of $H$ upto a point which is further down the
line $D$. By repeating this calculation with bigger and bigger cluster
sizes $n$, we can show that (5) is the ground state down to about
$\delta =0.136$. At that value of $\delta$, the cluster size $n$ is
as large as the largest system sizes $N$ that we we have studied.
Hence the argument that (5) is the ground state cannot be continued any
further. Below $\delta =0.136$, we have the `hole' where computations
are difficult. Since the segment of the straight line from the point
$(0,1)$ upto the `hole' has an exactly known ground state with an
extremely short correlation
length (essentially, one site), and since there is a cross-over from a Neel
phase to a spiral phase across the line, we may call it a disorder line
just as in the spin-$1/2$ case [10].

To summarize, we have studied a two-dimensional phase diagram for the
ground state of an isotropic antiferromagnetic spin-$1$ chain. It is
considerably more complicated than the corresponding spin-$1/2$ chain [10]
with surprising features like an almost gapless point inside the spiral
phase. We have suggested that this point is close to a critical point of
a particular kind. It would be interesting to establish this more
definitively. In any case, our results show that frustrated spin chains
with small values of $S$ may exhibit features not anticipated from
large $S$ field theories.

We thank B. Sriram Shastry for stimulating discussions.

\vfill
\eject

\line{\bf References \hfill}
\vskip .2in

\noindent
\item{1.}{T. Tonegawa and I. Harada, J. Phys. Soc. Jpn. {\bf 56},
2153 (1987), and references therein.}

\noindent
\item{2.}{S. R. White and D. A. Huse, Phys. Rev. B {\bf 48}, 3844
(1993).}

\noindent
\item{3.}{Y. Kato and A. Tanaka, J. Phys. Soc. Jpn. {\bf 63}, 1277
(1994).}

\noindent
\item{4.}{R. R. P. Singh and M. P. Gelfand, Phys. Rev. Lett. {\bf 61},
2133 (1988); D. Guo, T. Kennedy and S. Mazumdar, Phys. Rev. B {\bf 41},
9592 (1990).}

\noindent
\item{5.}{F. D. M. Haldane, Phys. Lett. {\bf 93A}, 464 (1983); Phys.
Rev. Lett. {\bf 50} 1153 (1983).}

\noindent
\item{6.}{I. Affleck, in {\it Fields, Strings and Critical Phenomena},
eds. E. Brezin and J. Zinn-Justin (North-Holland, Amsterdam, 1989);
I. Affleck and F.D.M. Haldane, Phys. Rev. B {\bf 36}, 5291 (1897).}

\noindent
\item{7.}{We use the word "phase" only for convenience to distinguish
between regions with different modulations of the two-spin correlation
function. Our model actually has no phase transition from Neel to
spiral even at zero temperature.}

\noindent
\item{8.}{S. Rao and D. Sen, Nucl. Phys. B {\bf 424}, 547 (1994).}

\noindent
\item{9.}{D. Allen and D. Senechal, Phys. Rev. B {\bf 51}, 6394 (1995).}

\noindent
\item{10.}{R. Chitra, S. Pati, H. R. Krishnamurthy, D. Sen and S.
Ramasesha, preprint no. cond-mat/9412016, to appear in Phys. Rev. B.}

\noindent
\item{11.}{S. R. White, Phys. Rev. Lett. {\bf {69}}, 2863 (1992); Phys.
Rev. B {\bf 48} ,10345 (1993).}

\noindent
\item{12.}{B. Sutherland, Phys. Rev. B {\bf 12}, 3795 (1975).}

\noindent
\item{13.}{I. Affleck, Nucl. Phys. B {\bf 265}, 409 (1986).}

\noindent
\item{14.}{At the $SU(3)$ symmetric critical point, the two-spin
correlation should asymptotically decay as the $4/3$ power of the
distance [13]. We tried to verify this but the finite correlation
length prevented us from obtaining an accurate estimate of the power.}

\noindent
\item{15.}{S. Rao and D. Sen, Phys. Rev. B {\bf 48}, 12763 (1993).}

\noindent
\item{16.}{M. den Nijs and K. Rommelse, Phys. Rev. B {\bf 40}, 4709
(1989); H. Tasaki, Phys. Rev. Lett. {\bf 66}, 798 (1991).}

\noindent
\item{17.}{T. Kennedy, J. Phys. Condens. Matter. {\bf 2}, 5737 (1990);
I. Affleck, T. Kennedy, E. H. Lieb and H. Tasaki, Comm. Math. Phys.
{\bf 115}, 477 (1988).}

\noindent
\item{18.}{B. S. Shastry and B. Sutherland, Phys. Rev. Lett. {\bf 47},
964 (1981).}

\vfill
\eject

\line{\bf Figure Captions \hfill}
\vskip .2in

\noindent
\item{1.}{Phase diagram for spin-$1$ in the ($J_2 , \delta$) plane. The
solid line $A$ extending from $(0.25,0)$ upto the cross is gapless. The
rest of the diagram is gapped. On the dotted lines $B$, the gap is finite.
The dashed line $C$ close to $(0.725,0)$ is almost gapless. The ground
state for an open chain has a
four-fold degeneracy below the curve $ABC$, while it is unique above
$ABC$. The straight line $D$ satisfying
$2 J_2 + \delta = 1$ extends from $(0,1)$ to about $(0.432,0.136)$.
Below $ABC$, there is a line $E$ which goes down
to about $(0.39,0)$. Across $D$ and $E$, the position of the peak in
the structure factor decreases from $\pi$ (Neel) to less than $\pi$
(spiral).}

\noindent
\item{2.}{Dependence of the gap on $J_2$ for $\delta =0$.}

\noindent
\item{3.}{Structure factor $S(q)$ versus $q$ for $J_2 =0.71$, $0.72$,
$0.725$ and $0.735$ at $\delta =0$.}

\end